# Microfabricated electromagnetic filters for millikelvin experiments

Hélène le Sueur and Philippe Joyez

Service de Physique de l'Etat Condensé (CNRS URA 2464)

CEA Saclay

91191 Gif sur Yvette Cedex, France

**Abstract.** In this article we report on the design, fabrication, and tests of microfabricated broadband filters suitable for proper electromagnetic thermalization of electrical lines connected to sensitive quantum electronics experiments performed at dilution fridge temperatures. Compared to previous such miniature filters, the new design improves on performance and reliability. These filters can be packed in space-saving cases with either single or multicontact connector. Measured performance in the accessible range compares well to simulations. We use these simulations to discuss the effectiveness of these filters for electromagnetic thermalization at 30 mK.

**INTRODUCTION.** A number of quantum electrical experiments need to be operated at millikelvin temperatures for optimum performance. One can think, for instance, of solid-state qubits or of tunnel spectroscopy experiments (in which energy resolution is directly proportional to the temperature). If cooling a sample down to the millikelvin range is rather easy, ensuring that an experiment is performed at full thermal equilibrium is not. The reason is well known: leads connecting the sample to room temperature apparatus feed in electromagnetic noise, which may directly couple with the device under test (e.g., in the case of a qubit) and/or heat up the electrons of a small device above the temperature of the crystalline network. Indeed, many experimental results in the field of low temperature quantum electronics can only be interpreted with an electronic temperature greater than that of the refrigerator. The noise generated by room temperature instruments is at the minimum the Johnson thermal noise (the electrical equivalent of Planck's blackbody radiation) extending in frequency up to 300 K.$k_B$/h ~ 6 THz, in the infrared frequency range. One, thus, needs to insert in the leads connected to the sample filters able to absorb this very wide band noise and to reemit a thermal spectrum at the experiment temperature which corresponds to a bandwidth of a fraction of 1 Ghz. This problem has been investigated theoretically, and the stringent requirements this imposes on noise attenuation have been evaluated for operating Coulomb blockade devices [1,2].

Unfortunately, there exists no ready-made commercial filters with such wide stopband and sufficient attenuation. This is why in the last 25 years many custom filter designs have been proposed in order to effectively thermalize experiments at millikelvin temperatures. Most solutions rely on the use of shielded lines (usually coaxial lines) and shielded distributed *RC* filters. These filters being dissipative, they must themselves be placed at the lowest temperature so that their own thermal noise is at equilibrium. A popular solution (easy to make and use) is to combine both the line and the filter by using a lossy coax such as Thermocoax® [3,4]. One drawback with this approach is that the length required for a low enough cutoff frequency is often of the order of meters, which is impractical to install in many dilution fridges where space is often scarce. Moreover, when such a line is running between two or more different temperatures, estimating the resulting output noise temperature is not easy. Other popular filters are powder filters based on skin effect in metallic grains [5]. These are very effective filters, but once again they tend to be rather large and their performance is not reliably predictable at design time.

Some ten years ago, with the need for space-saving filters for Coulomb blockade experiments, one of us was involved in the design and fabrication of miniature filters, as reported in Ref. [2]. These filters were based on microfabricated chips made by optical lithography which consisted of a resistive meander line sandwiched between two insulating layers and ground planes, implementing a distributed *RC* filter. These filters had a predictable performance in the 0-20 GHz range, which could be tailored at design time. They had several weaknesses, though: (i) their very high frequency attenuation was intrinsically limited by residual coupling between adjacent meander lines and (ii) it was further degraded by shortcomings in the casing design. Finally, (iii) these filters had reliability issues upon thermal cycling attributed to differential contraction of the various materials involved in the filter assembly. Critically reviewing these problems, we came up with a new design of microfabricated filters which corrects all these problems. We now consistently obtain much better stop-band attenuation together with other desirable characteristics: less dc resistance, less total capacitance, better aging, and high-voltage capability. In this article we describe this design, along with the fabrication, packing, and performance of the corresponding filters.

**DESIGN.** When designing a filter such as we need, one can imagine that very high frequency electromagnetic waves (hereafter microwaves for short, but extending, in fact, up to infrared) will sneak into the least metallic hole, slot, or crack and propagate through vacuum (or dielectric) much like a fluid would: If there is a dielectric path between input and output, microwaves will find their way. Thus, designing an efficient filter is much like making the device leak-tight to fluids. The key step towards electromagnetic "hermeticity" in our new design is the use of a brass substrate for the microfabricated filter chip in place of the more standard silicon substrate used in the old design. This way, not only do we suppress the dielectric substrate which is difficult to seal perfectly and which behaves as a wave guide, but we also suppress differential thermal contraction of the filter chip with respect to the brass case, which ensures that no cracks will appear upon thermal cycling. The design of the filter chip itself is shown in Fig. 1. It consists of a resistive meander line connecting the input pad to the output pad. The meander line alternates narrow parts and large pads. This provides a steeper cutoff than a meander with uniform width having the same total resistance and capacitance, due to reflections on impedance discontinuities. The meander line is completely surrounded by an insulating material and a ground sheath, which provides decoupling of adjacent arms of the meander at high frequency. Apart from the different sections, it is like a miniature distributed resistive coaxial cable. Its behavior is thus somewhat similar to that of Thermocoax-type filters, but its waveguide modes are rejected at higher frequencies due to the smaller cross-sections of the dielectric.

**FABRICATION.** Fabrication starts from 0.4 mm-thick brass sheet, out of which we cut 75 mm diameter wafers. Each wafer is mechanically polished down to a 1/4 μm grain size to yield a fair optical polish. Then we deposit the bottom insulating layer using spin coating. This layer consists of Cyclotene 4024 bisbenzocyclobutene-based (BCB) negative photoresist from Dow Chemicals. For our requirements we chose a spin speed of 6000 rpm, thus obtaining a resist thickness of 2.5 μm. By varying the dilution of the photoresist or/and the spin speed we can tune this thickness reliably down to 500 nm and change the filter capacitance accordingly. The resist is then patterned using UV photolithography. The wafer is subsequently immersion-developed and submitted to a moderate cure at 210 °C under inert atmosphere to convert about 75% of the resist into the final polymer. This hardens the layer, which can then handle further microfabrication steps, and enhances subsequent layer adhesion. In this step, the BCB loses ~10% thickness. Then, following the manufacturer's recommendation, the wafer is submitted to a $SF_6/O_2$ reactive ion etching plasma to eliminate resist residues and enhance subsequent layer adhesion on BCB. This plasma etching further reduces the resist thickness by about 100 nm. Next, we spin on a 1.3 μm thick layer of Shipley S1813 positive photoresist. The meander pattern is then exposed, aligned onto the previous insulating layer. Prior to developing, the wafer is immersed for 30 s at 30°C in dichlorobenzene to obtain after developing an overhanging resist profile needed for a good lift-off. The substrate is then placed in a Joule evaporator and we deposit a 5 nm-thick titanium adhesion layer followed by a 100 - 400 nm thick copper-gold alloy at 50% atomic content with a resistivity of $\sim 1.3 \times 10^{-7}\,\Omega$m to form the resistive meander. The thickness of the layer is chosen to obtain the desired total resistance of the meander and filter characteristics. As copper and gold evaporate at different rates, we prepare a fresh load at each pump down and evaporate it completely to obtain a reproducible alloy. After deposition the resist is lifted-off in acetone. In a third lithographic step, we pattern a second BCB layer to cover the meander in the central part of the chip while leaving uncovered the substrate between arms of the meander and the connecting pads of the meander (see Fig. 1). The BCB is finally hard cured for 1 h at 250 °C under inert atmosphere to achieve 98% polymer conversion and bring it to nominal electrical and mechanical characteristics. In the last lithographic step, another layer of Shipley S1813 is used to evaporate a ~1 μm-thick gold sheath on top of the meander to terminate encapsulation and electrical shielding of the meander in the central part. Finally, the brass wafer is diced into 3x8 mm$^2$ chips for individual filters. For multiline filters, several adjacent filters are left attached together.

**ASSEMBLY.** Filter chips are then mounted in cases. In Fig. 2 we show the design of a case for a single filter equipped with soft-soldered standard SMC coaxial connectors. The goal is to mount the filter chip in-between two perfectly separated chambers. The two chambers are separated by a small metallic barrier running across the meander. Perfect electromagnetic separation of the chambers is reached by filling all machining gaps with silver epoxy. In each chamber we establish a connection from the connector pin to the filter pad using thin wire and silver epoxy. Finally, we close the cover applying here also silver epoxy to make it electromagnetically leak-tight. Following the same design, other cases with different connectors can be made. In particular, double filters for shielded twisted pair lines are assembled in cases fitted with miniature triaxial connectors (Fischer series 101), and up to 25 filters can be packed in a 34.5 x 30.1 x 13 mm$^3$ case equipped with two 25 contact shielded micro-subD connectors.

**TESTING.** Figure 3 shows the measured attenuation of a single filter equipped with SMC coaxial connectors. These measurements were performed using a 65 GHz, two 50 Ω port vector network analyzer (Anritsu 37297D). The $S_{21}$ forward transmission data show that the filter has a steep cutoff reaching an attenuation greater than the ~ -100 dB noise floor of the analyzer at 4.5 GHz and remains below the noise floor up to 65 GHz. Attenuation measurements carried out at 4.2 K (data not shown) show little change in the response of the filter, essentially due to a ~20% reduction of the resistivity of the alloy we use. However, due to losses in the cables running from the network analyzer into the liquid helium Dewar, the sensitivity of this low temperature measurement is much lower than that of the room temperature measurements shown in Fig. 3, and thus it is less suited to test for proper attenuation performance. Repeated thermal cycling of the filter showed no measurable change in properties. Since one application for these filters in our laboratory will be to filter high voltage lines connected to piezoelements in a low temperature scanning probe microscope, we further submit all assembled filters to a 30s, 400 V test between the meander line and ground. A significant fraction of the filters do not pass this test, developing a permanent short to ground. According to the specifications, the BCB breakdown field is greater than 300 V/µm. Given the BCB thicknesses used, 400 V should not exceed the breakdown voltage of the dielectric. We think the lower breakdown voltage observed in many filters is due to the field enhancement occurring on sharp defects. The overall yield after assembly and high voltage test is of the order of 50%, which could likely be improved by a better control of process parameters and cleanliness. Note that given their rather low resistance, these filters are acceptable on lines driving ~10 nF piezoelements used for inertial stick-slip motion, where a large instantaneous current is needed.

**SIMULATIONS.** We now compare the measurements shown in Fig. 3 to the expected behavior of this filter. For this we have modeled the full filter chip using a 2.5-D planar electromagnetic simulator (Sonnet). Given the feature sizes and aspect ratios used, approximations made in such a simulation are valid throughout the whole frequency range we consider [6]. The data input in the geometrical model were the nominal dimensions of the meander line and the measured thickness of both BCB layers (1.6 µm for the bottom layer and 2.2 µm for the top layer) obtained using a thin film profile meter. The sheathing of the meander is modeled using vias between metallic layers above and below the dielectric layers. Electrical data for the BCB are taken from the manufacturer : $\varepsilon_r$ = 2.5, the relative dielectric constant, and a loss tangent of $2 \times 10^{-3}$. These inputs yield an 81 pF capacitance to ground of the filter at low frequency, in remarkable agreement with the 80 pF measured value. The conductivity of the meander is adjusted to reproduce the 77 Ω dc resistance of the filter. Given these inputs, the calculated forward transmission also shown in Fig. 3 is in very good agreement with the measurements. The small discrepancy showing up above 2 GHz might be due to the fact that we do not take into account the circuit outside the filter chip. For instance, the small wires connecting the chip to the connectors introduce some inductance in series with the filter, which increases attenuation as frequency rises. The connectors also introduce losses in the transmission above 10 GHz. These simulations also show the superiority of the present design compared to the former one [2]: if we remove isolation between adjacent arms of the meander, the filter transmission remains between -85 and -95 dB in the 5 - 100 GHz range (see inset of Fig. 3).

Given the good agreement obtained, one can wish to push this simulation up to higher frequencies to get insight on the behavior of the filter in ranges where it would be extremely difficult to measure. However, it turns out that the full chip simulation remains at the simulator noise floor (between -150 and -170 dB, depending on the frequency [7]) above 12 GHz, and furthermore, becomes too large to be performed on an average workstation for frequencies larger than ~100 GHz. In order to go beyond these limitations one needs to recourse to subdividing the filter into smaller parts and to cascading the resulting $S$ matrices. Note that this assumes that the subdivided parts are connected exclusively through their connecting ports (i.e., no coupling through the sheathing, which is what one expects from a bulk metallic shield). Doing so, we can push the simulation up to 6 THz without excessive complications (see inset of Fig. 3). We find that above 50 GHz oscillating features appear with a 95 GHz period, corresponding to standing waves in the 1 mm-long features of the meanders, and a second one at ~ 400 GHz corresponding to a standing wave in the width of a large pad in the meander. To sketch the filter behavior, we can say that after the initial cutoff, the transmission curve forms a quasiplateau between roughly 20 and 400 GHz with a maximum transmission of ~ -220 dB and subsequently falls rapidly again with an overall $\exp(-\sqrt{f})$ dependence, due to skin effect losses [8]. Note that these simulations do not take into account a number of effects which might be important at very high frequency such as the increase of the loss tangent of the dielectric or the surface roughness of the conductors, which likely would further increase filter attenuation.

**DISCUSSION.** Based on these simulations, we can evaluate the effectiveness of these filters for experiment thermalization. To do so, we calculate the effective temperature seen by a device connected to a line comprising a 50 Ω source at 300 K and either a single filter anchored at 30 mK or two filters respectively placed at 4.2 K and 30 mK. This frequency-dependent effective temperature is the temperature at which an impedance equal to the output impedance of the line should be to have the same noise level as that reaching the device connected to the line at 30 mK at a given frequency. The equivalent noise source at the output of the line can be evaluated exactly by adding properly the contributions coming from the various parts of the line situated at different temperatures weighted with the corresponding attenuation provided by the setup [9]. The result is displayed in Fig. 4 and shows that effective 30 mK thermalization is reached in the 1-30 GHz range when using two filters in series [10]. At higher frequencies, the photon flux at the output of the line is larger than that of a 30 mK impedance but remains, in absolute, so low that its probability to spoil an actual experiment is negligible. On the lower end of the spectrum, the effective temperature is also above equilibrium but corresponds to photons present in the thermal spectrum at 30 mK. Further reduction of noise in this part of the spectrum may be desired, in particular, to protect the experiment against electromagnetic interferences (EMIs) in the laboratory ambient. In this frequency range, this can be done using carefully assembled discrete elements. In this goal, the connecting pad of the filter chip were made to accommodate 0805-size surface mounted components such as readily available lossy inductors sold for EMI filtering with good attenuation in the 0.1-1 GHz range.

At low temperature, Joule effect in such a resistive filter can easily heat the electrons of the filter out of equilibrium, due to the poor electron-phonon coupling [11]. Therefore, care should be taken not to drive excessive currents through these filters. For a given allowed relative increase $\eta$ of the electronic temperature, we can evaluate at lowest order the maximum dc current allowed to flow in the filter as $i_{max}=\sqrt{5\sigma V T_{ph}^5(\eta-1)/R}$, where $\sigma \sim 1$ nW/µm$^3$/K$^5$ is the coupling constant, $V$ is the volume of the resistive material volume where electron-phonon coupling takes place, $T_{ph}$ is the phonon temperature, and $R$ is the resistance of the filter. To fix ideas, if one takes the whole volume of resistive material of the filter measured above, $V = 1.5\ 10^6$ µm$^3$, and a 5% increase in the electron temperature ($\eta = 1.05$) above $T_{ph} = 30$ mK, one gets a maximum allowed dc current $i_{max}$ = 0.34 µA.

**POSSIBLE IMPROVEMENTS.** As mentioned above, the resistivity of the copper-gold alloy we use depends on temperature. This alloy was chosen mainly for its ease of use in our Joule evaporator as it yields a relatively low-stress film. This alloy could be replaced by another type of resistive alloy less dependent on temperature, provided maybe another deposition technique was used. Possible choices could then be Ni-Cr, Cu-Ni or Cu-Mn.

**ACKNOWLEDGEMENTS.** The authors gratefully thank P.F. Orfila, P. Senat, D. Vion, and D. Esteve for their help and support throughout this development, and N. Feltin, B. Steck, and P. Bertet for their interest and involvement in this project.  They acknowledge useful discussions with H. Courtois and D.C. Glattli. This work was supported in part by the French Research Ministry

**FIGURES**

Figure 1.

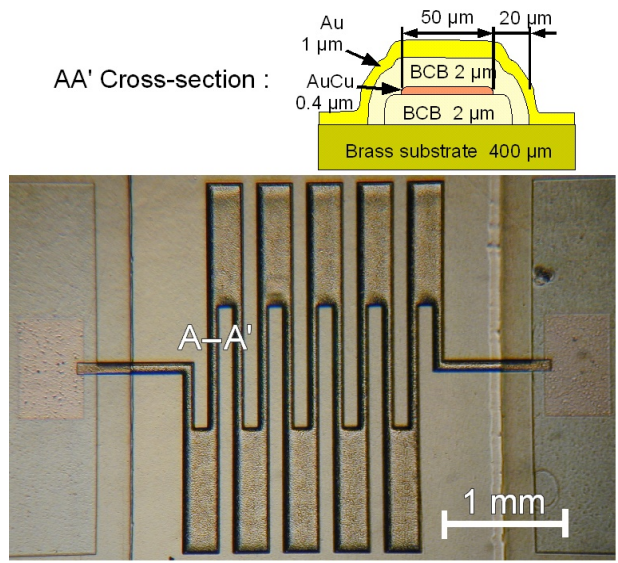

Picture of a microfabricated filter chip and crosssection of a meander arm with materials and respective thickness (note different vertical and horizontal scales).

**Figure 2.**

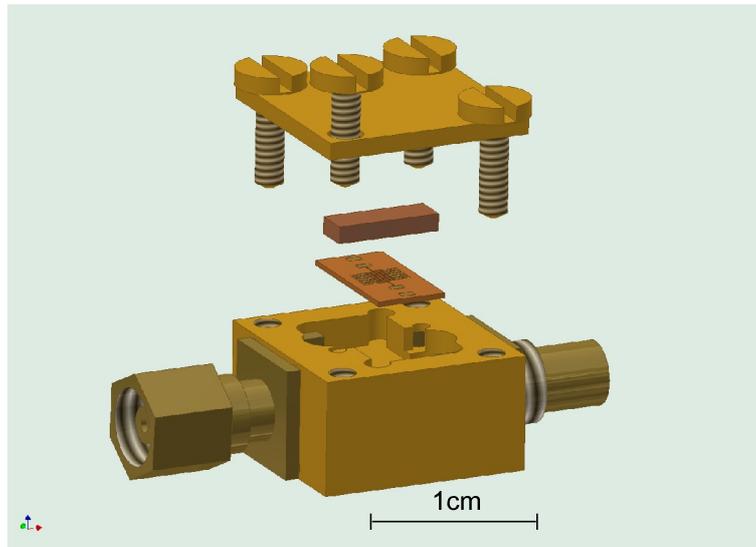

Exploded view of a single filter case equipped with SMC coaxial connectors. Connections between connector pins and filter pads are made with thin insulated wire and silver epoxy (not shown). All machining gaps and joints are filled with silver epoxy to achieve perfect electromagnetic separation between input and output connector chambers. Screws can be used to thermally anchor the case.

**Figure 3.**

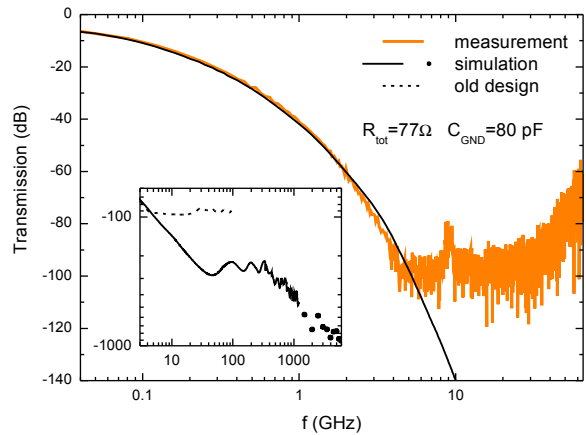

In the main panel, transmission of an assembled filter having a dc resistance of 77 Ω and capacitance to ground of 80 pF. The data were measured using a microwave network analyzer up to 65 GHz. Above ~ 4.5 GHz the noise floor of the analyzer is reached (the peak at 9 GHz is also noise and is due to a calibration artifact). We also plot the results of a microwave simulation with no adjustable parameters (see text). In the inset, results of the microwave simulation at higher frequency and predictions for the transmission of the same filter if the meander arms were not screened, as in Ref. [2].

**Figure 4.**

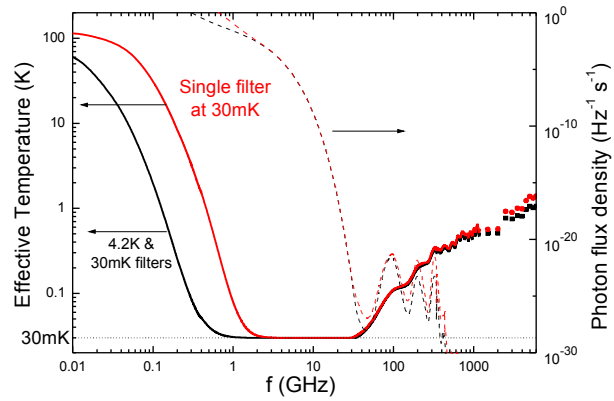

Calculated effective temperature (full lines, left axis) seen by a device connected to a line comprised of a 50 Ω source at 300 K and either a single filter (red curves) anchored at 30 mK or two filters (black curves) respectively placed at 4.2 K and 30 mK and photon flux density (dashed lines, right axis) at the output of this line on a 50 Ω load.